\documentclass[conference]{IEEEtran}

\usepackage{graphicx}
\usepackage{color}
\usepackage{algorithm}
\usepackage{algorithmicx}
\usepackage[lined,boxed,commentsnumbered]{algorithm2e}
\usepackage{epstopdf}

\begin{document}

\title{An Adaptative Multi-GPU based Branch-and-Bound. A Case Study: the Flow-Shop Scheduling Problem}

\author{	\IEEEauthorblockN{I. Chakroun, N. Melab }

\IEEEauthorblockA{Universit\'e Lille~1, LIFL/UMR CNRS 8022 - INRIA Lille Nord Europe\\
59655 - Villeneuve d'Ascq cedex - France \\
Email: \{imen.chakroun, nouredine.melab\}@lifl.fr}
}

\maketitle

\begin{abstract}

Solving exactly Combinatorial Optimization Problems
(COPs) using a Branch-and-Bound (B\&B) algorithm requires a
huge amount of computational resources. Therefore, we recently investigated 
designing B\&B algorithms on top of graphics processing units (GPUs) using a parallel bounding model.
The proposed model assumes parallelizing the evaluation of the lower bounds on pools of sub-problems.
The results demonstrated that the size of the evaluated pool has a significant impact on the performance of B\&B
and that it depends strongly on the problem instance being solved. In this paper, we design an adaptative parallel B\&B algorithm for solving 
permutation-based combinatorial optimization problems such as FSP (Flow-shop Scheduling Problem) on GPU accelerators.
To do so, we propose a dynamic heuristic for parameter auto-tuning at runtime.  
Another challenge of this pioneering work \footnote{To the best of our knowledge, 
our work is the first implementation of an adaptative Branch and Bound on multi-GPUs platforms.} is to exploit larger degrees of parallelism
by using the combined computational power of multiple
GPU devices. The approach has been applied to the
permutation flow-shop problem. Extensive experiments have been carried out on well-known FSP benchmarks using an 
Nvidia Tesla S1070 Computing System equipped with two Tesla T10 GPUs. Compared to a CPU-based execution, 
accelerations up to $\times$105  are achieved for large problem instances. \\

\end{abstract}


\section{Introduction}

Solving to optimality large size combinatorial optimization problems (COPs) 
\footnote{An optimization problem consists in minimizing or maximizing a cost function.
Without loss of generality, in this paper the minimization case is considered.} using a
Branch and Bound algorithm (B\&B) is CPU time-consuming. Although B\&B allows 
to reduce considerably the exploration time using a bounding mechanism,
often only small or moderately-sized instances can be practically
solved. Therefore, over the last decades, parallel computing 
has been revealed as an attractive way to deal with 
larger instances of COPs. However, while many contributions have been proposed 
for parallel B\&B methods using Massively Parallel Processors \cite{Allen_1997}, Networks or 
Clusters of Workstations \cite{Quinn_1990} and SMP machines \cite{Casadoa_2008}, very few 
contributions have been proposed for designing B\&B algorithms on Graphical Processing Units (GPUs) \cite{Chakroun_2011}, \cite{Boukedjar_2011}. 
For years, the use of graphics processors was dedicated to graphics applications. Driven
by the demand for high-definition 3D graphics on personal computers, GPUs have evolved
into a highly parallel, multithreaded and many-core environment. 
Their utilization has recently been extended to other application domains such
as scientific computing \cite{Kurzak_10}. 


Most of existing parallel B\&B algorithms, such as the above ones, are based on the parallel exploration 
of the search tree. Such parallel model is not suited to GPUs because the explored search tree is highly 
irregular. In our work \cite{Chakroun_2011}, we proposed a pioneering 
investigation of using a parallel bounding model for designing B\&B algorithms over GPUs. The proposed model 
assumes parallelizing the evaluation of the lower bounds on pools of sub-problems.
The experimental results show that significant accelerations can be obtained 
especially for large problem instances and large pools of subproblems.
Results demonstrate also that the size of the evaluated pool has an important impact on the performance of the B\&B
and that it depends strongly on the problem instance being solved. It is thus hard to fix it a priori and so has 
to be tuned dynamically depending on the problem instance being tackled.

In this paper, we design an adaptative parallel B\&B algorithm for solving 
permutation-based combinatorial optimization problems such as FSP (Flow-shop Scheduling Problem) on GPU accelerators. 
The idea is to dynamically tune the size of the pool being off-loaded to the GPU taking into consideration both the characteristics
of the used device and the problem instance being tackled. Another challenge of this work is to exploit larger degrees of parallelism
by utilizing multiple GPUs. Indeed, execution on parallel GPUs is promising since applications
that are best suited to run on GPUs inherently have large amounts of parallelism. Using multiple GPUs avoids
also dealing with the limitations of devices, like memory resources, by exploiting the combined resources of
multiple boards. 

The remainder of the paper is organized as follows: Section~\ref{BB} presents the B\&B algorithm and the permutation Flow-shop Scheduling Problem.
In Section~\ref{GPUBB}, we describe our adaptative GPU-based proposed approach for B\&B. 
In Section~\ref{MULTIGPU}, we describe our methodology for using multiple GPUs.
In Section~\ref{Experiments}, we report experimental results demonstrating the efficiency of our approach. 
Finally, some conclusions and perspectives of this work are drawn in Section~\ref{Conclusion}.

\section{B\&B for the Permutation Flow-shop Scheduling problem}
\label{BB}

\subsection{B\&B algorithms}
\label{BBAlgo}

Branch-and-Bound (B\&B) algorithms are well-known exact methods for solving to 
optimality combinatorial optimization problems. They are based on an implicit enumeration
of all the solutions of the considered problem.
The search space is explored by dynamically building a tree whose 
root node designates the original problem. The construction of the B\&B 
tree and its exploration are performed using four operators: {\it branching}, 
{\it bounding}, {\it selection} and {\it elimination}. The algorithm proceeds 
in several iterations during which the best solution found so far is progressively improved. 
The generated and not yet examined sub-problems are kept into a list initialized to the original problem.
At each iteration, a sub-problem is selected from this list according to some defined strategy
(depth-first, best-first,$\ldots$), using the {\it selection operator}.
Then, a {\it branching operator} is applied on the selected sub-problem,
subdividing its solution space into
two or more subspaces to be investigated in a subsequent
iteration. For each one of the generated sub-problems, the {\it bounding operator} 
calculates a lower bound that is compared to the upper-bound. 
Each sub-problem having a greater bound than
the upper-bound, the cost of the best solution found so far,
is pruned using the {\it elimination operator}.

Thanks to the bounding operator, B\&B allows to reduce considerably the computation time needed
to explore the whole solution space. However, the exploration time remains significant and parallel
processing is thus required. In~\cite{BGendron_et_al_94}, three parallel models are identified for B\&B algorithms: parallel 
application of the operators on the generated sub-problems (Type~1), parallel building and exploration of a B\&B tree (Type~2), and
parallel (cooperative or independent) building and exploration of several B\&B trees (Type~3). We have already 
rethinked these parallel approaches for large-scale computational grids~\cite{Djamai_2011} using Type~2 parallel model. 
Grid computing provides an impressive computing power to solve challenging instances in combinatorial 
optimization~\cite{Mezmaz_2007}. However, computational grids providing a huge amount of resources are not easily available 
and accessible for any user. Recently, Graphics Processing Units (GPU accelerators) have emerged as a new popular support for 
massively parallel computing. GPUs are high-performance many-core processors capable of very high computation 
and data throughput. Such resources are also energy-efficient and unlike grids they are highly 
available every where. In the following, we use the Type~1 parallel model on GPU for solving 
Flow-Shop problems.

\subsection{The permutation Flow-shop Scheduling problem}

\begin{figure}
  \begin{center}
\includegraphics[width=9cm]{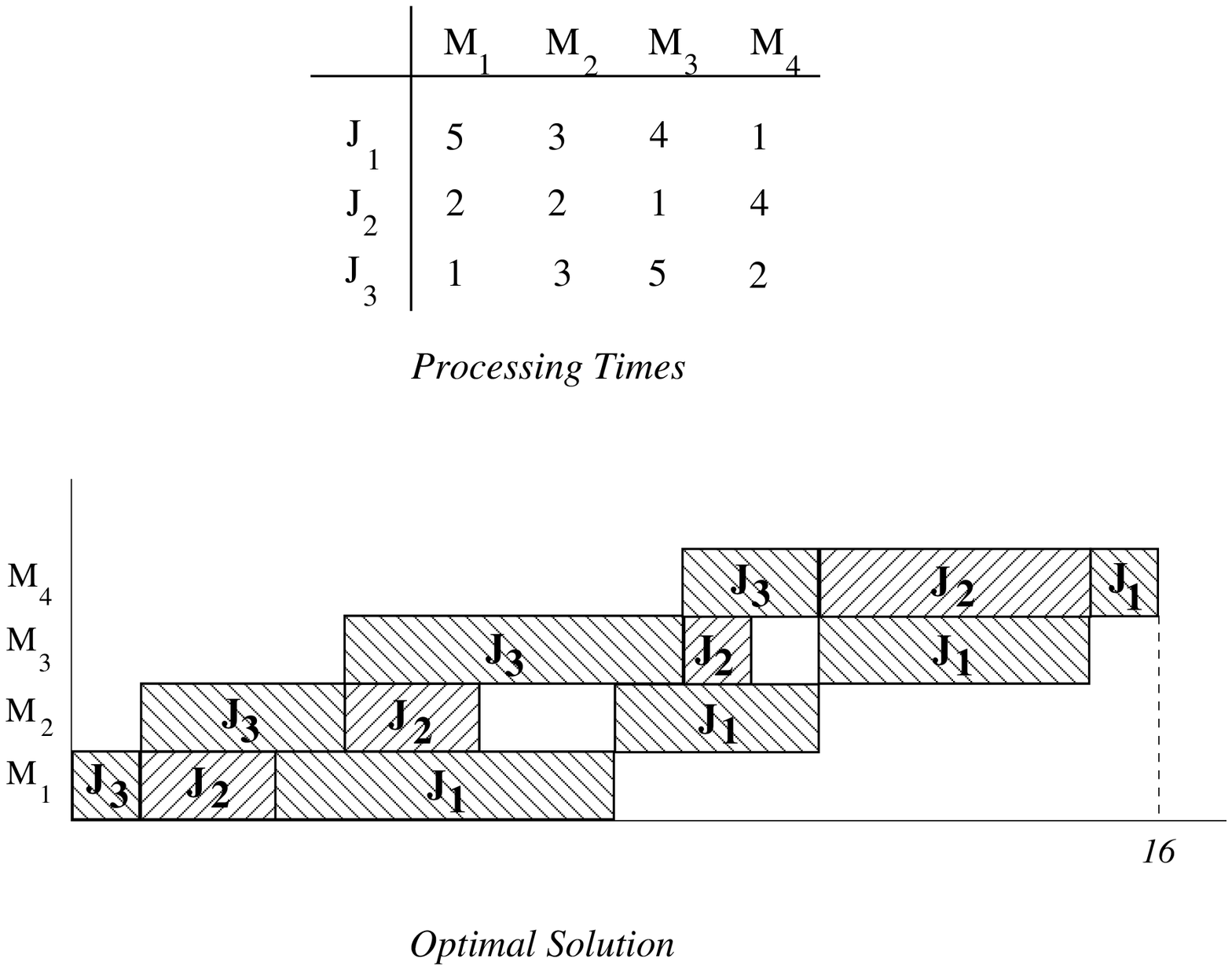}%
  \caption{Illustration of a permutation FSP with $n=3$ and $m=4$. The table reports 
the processing times of the jobs on the machines. The Gantt diagram shows the optimal solution to the problem instance.}
\label{FSPExample}
\end{center}
\end{figure}

The general FSP can be formulated as follows~\cite{JKLenstra_et_al_78}. FSP consists in scheduling a pool of $n$ jobs on a set of $m$ 
machines such that each of the jobs $J_1$, $J_2$, \ldots, $J_n$ has to be processed on the machines $M_1$, $M_2$, \ldots, $M_m$ 
in that order. Job $J_i$ (i = 1, 2, \ldots, n) consists therefore of a sequence of $m$ operations $O_{i1}$, $O_{i2}$,  
\ldots $O_{im}$; $O_{ik}$ being the processing of $J_i$ on $M_k$ during an uninterrupted processing time $p_{ik}$. $M_k$ (k = 1, 2, \ldots, m) 
can handle at most one job at a time. The objective is to find a processing order on each $M_k$ such that the time required to complete all 
jobs is minimized. If the problem is restricted to the minimization over all permutation schedules, meaning with the same processing 
order on each machine, the resulting problem is called the permutation Flow-Shop problem, which is the focus of this work. Figure~\ref{FSPExample} 
shows an example of an FSP instance (with $n=3$ and $m=4$) and its associated optimal solution.

In the B\&B applied to the the FSP, the node number $i$ in the search tree represents 
the sub-problem in which job $J_i$ is scheduled first on all machines.
The decomposition of this problem generates $n$ sons, each of them designates a sub-problem. 
The recursive application of the decomposition operator on the generated sub-problems allows 
to develop the search tree. The number of potential schedules (permutations) is $n!$, which is extremely large 
for large problem instances such as $200 \times 20$
($200!$ schedules!) Taillard's ones~\cite{Taillard_Bench}. To speedup the exploration of such large search
trees, two major powerful ways are used. The first way consists in using an efficient bounding
operator. Applied to a sub-problem, such operator associates a value
to its corresponding tree node using a lower bound function. If this lower bound value is
greater than the cost of the best schedule found so far (upper bound), the sub-problem is not
decomposed and its tree node is pruned. The second
way is to use massively parallel computing based on the three
parallelism types presented in the section \ref{BBAlgo}. We remind
that the focus of this paper is only on {\it Type~1} i.e. the
parallel evaluation of the lower bound on a pool of sub-problems.

In the following, we present a new auto-adaptative GPU-based approach for the parallel evaluation of the 
lower bound in B\&B algorithms. 

\section{Our Adaptative GPU-based B\&B algorithm}
\label{GPUBB} 

The proposed approach is based on the GPGPU (CUDA or OpenCL) parallel paradigm. 
According to this paradigm, the programmer writes a serial program that calls parallel kernels.
The kernel is the core code that defines the computation to be performed by a
large number of threads. These threads are organized in collections called blocks that
can be assigned to a single multiprocessor and which execution is time-shared. 
A collection of all blocks in a single execution is called a grid. 

\begin{figure}[h!]
  \begin{center}
\includegraphics[width=9cm]{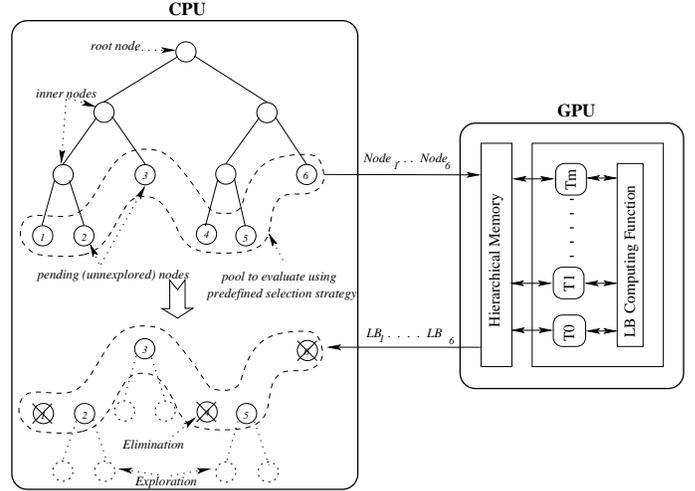}
\caption{The overall architecture of our GPU-accelerated branch-and-Bound algorithm. Our approach introduces 
two main adaptations compared to a traditional B\&B : selection of thousand of nodes and evaluation in parallel.}
\label{approach}
  \end{center}
\end{figure}

In our revisited GPU-based B\&B algorithm, the generation (elimination, selection and branching operations) of the sub-problems to be solved  
is performed on CPU and the evaluation of their lower bounds (bounding operation) is executed on the GPU device. As illustrated in Figure~\ref{approach}, 
the pool of sub-problems generated on CPU is off-loaded to the GPU device to be evaluated by a pool of threads. 
Each thread applies the lower bound function to one sub-problem. Once the evaluation is completed, the lower bound values are
returned back to the CPU to be used by the elimination operator. The process is iterated until the exploration is completed and the optimal solution is found.

One of the challenging concerns that must be considered to make efficient our GPU-based B\&B is supplying the device with a large pool of
subproblems. Indeed, in~\cite{Chakroun_2011}, experiments show that the proposed parallel bounding model 
is efficient only when large pools (thousands of sub-problems) are considered whatever 
the size of the FSP instances being tackled is. As a solution for the problem, we come up with a new selection strategy. 
Indeed, rather than selecting a single pending node as in traditional B\&B algorithms, 
our approach assume that a pool of pending nodes is selected from the search tree (see Figure~\ref{approach}). 
At each iteration of the algorithm, a pool of unexplored nodes is selected from the search tree according to their depth.
Deepest pending nodes are the first selected for being branched. As explained before, that pool of sub-problems, 
corresponding to the generated tree nodes and resulting 
from the branching operation, is off-loaded from CPU to GPU to be evaluated by blocks of threads. 

In our investigation proposed in~\cite{Chakroun_2011}, results also demonstrate
that the size of the pool to be off-loaded to the GPU has an important impact on the 
performance of the algorithm. We have also noticed that this parameter depends strongly
on the problem instance being solved. It is thus hard to be fixed a priori and so has to be 
tuned dynamically depending on the problem. For dealing with this issue, we propose an empirical heuristic for parameters auto-tuning at runtime. 
Algorithm \ref{auto} gives the general template for this heuristic.
The main idea of this approach is to send the pending sub-problems using different-sized ``waves'' to the GPU device 
during the first iterations of the B\&B algorithm. Regularly, we compute the efficiency of the used
pool and then double the size of the pool to off-load to the GPU. After a fixed number of trials, the better efficiency overall selected 
configurations is used for the remaining iterations of the algorithm. 
As explained above, the pool of sub-problems off-loaded to the GPU is evaluated by a pool of threads where each thread applies the lower bound function to one sub-problem. 
Consequently, the number of sub-problems to evaluate in parallel strongly depends of the total number of threads that would be triggered on the GPU.
Actually, tunning the size of the "wave" to submit to the GPU is equivalent to adjusting the number of the threads to run in parallel.

Our heuristic first identifies the characteristics of the used hardware. Thanks to this property, the algorithm becomes highly portable and could easily be run over
heterogeneous GPU architetures transparently to the user. The heuristic determines the maximum configuration that can be used, namely the maximum number of threads 
and blocks that can run in parallel over the GPU card. Indeed, in some cases, when a thread block allocates more registers than are available on a multiprocessor, 
the kernel execution fails since too many threads are requested. During all the tuning process, the number of threads per blocks is set using
the occupancy calculator tool provided by NVIDIA which allows the programmer to easily calculate 
the best thread block size based on register and shared memory usage of a kernel.
Regarding the number of blocks per grid, our primary concern when choosing this parameter was keeping the entire GPU busy. 
Indeed, the number of blocks in a grid should be larger than the number of multiprocessors so that all multiprocessors have at least one block to execute. 
Thus, we first initialize the number of blocks with the number of the multiporcessors detected on the device. 
This number is doubled repeatdly after a certain number of iterations (fixed experimentally) and until the number of threads
per blocks $\times$ the number of blocks doesn't exceed the maximum number of active threads allowed on the device. 

So far, our empirical search of the best efficiency is coarse-grained. Indeed, doubling the size in every step, and stopping when the efficiency is 
no longer improved, or when the limits of the GPU have been reached might founds an imprecise upper bound of the performance.  
For this reason and in order to make the tuning more thorough, we considered to also perform a binary search around the best pool size found so far.  
When the maximum number of active threads is reached, the iterative doubling proccess terminates and returns the best found configuration parameters.
The heuristic then computes a downwards and an upwards search around the best pool size found so far. The better efficiency overall selected 
configurations is used for the remaining iterations of the algorithm.

\begin{algorithm}[H]

\SetAlgoLined

\KwData{nb\_iterations;}

\KwResult{best\_number\_of\_threads}

max\_nb\_threads = Detect\_GPU\_Charateristics(); 

nb\_threads = Use\_Cuda\_Occupancy\_Calculator(); 

nb\_blocks := Get\_Number\_Of\_Multiporcessors();

\While{ not\_empty\_tree() }
{	
	\While{ pool\_size $\leq$ nb\_threads $\times$ nb\_blocks }
	{	
	    take\_sub\_problem(); 
	}

	Iteration pre-treatment on host side;\\
	Kernel evaluation on GPU;\\
	Iteration post-treatment on host side;
     
	\If{ ( iteration \% nb\_iterations = 0 ) and ( (nb\_threads $\times$ nb\_blocks) $\leq$ max\_nb\_threads) } 
 	{
	    \If{ Is\_best\_pool\_improved() }
	    {
	      best\_number\_of\_threads = nb\_threads $\times$ nb\_blocks ;
	    }
	    
	    nb\_blocks := nb\_blocks * 2 ;
	}
	\Else 
	{
	   Compute\_Binary\_Search\_Around\_Best\_Pool()   ;
	}
	   
	iteration := iteration + 1 ;
}

\caption{Dynamic parameter tuning heuristic}
\label{auto}
\end{algorithm}

\section{Reshaping the GPU-based B\&B algorithm for Multi-GPU architecture}
\label{MULTIGPU}

Nowadays, the trend in general-purpose computing on graphics processing units 
is to use multiple GPUs on a given system, much like using multiple cores on CPU-based
systems. In the following, we detail the changes we have made to our GPU-based B\&B algorithm 
presented in section \ref{GPUBB}. Our objective here is to consider the benefits of exploiting larger degrees of parallelism
by running our algorithm on multiple (parallel) GPUs.

The first step toward a multi-GPU design is to determine
how many GPUs will be used and how each GPU will be exploited.
In this work, our aim in using multi-GPUs is to speedup kernel execution rather than
utilizing each GPU differently (for example for evaluating different lower bound functions). 
Consequently, our concern here is to define a workload distribution between the used GPUs
in order to make all the available devices compute the same work in parallel without need of synchronization.
Since our approach ensures that the decomposed sub-problems are different and independent from each other
and since the used lower bound function is problem-dependent, we opted for simply splitting the pool of 
sub-problems among the selected GPUs. Each pool is then be evaluated in parallel and independently from other pools. 
However, after each GPU finishes computing the kernel function, the outputs from each device have to be merged to get final results.
The size of the pool to submit to each GPU is calculated using the proposed heuristic (see Section \ref{GPUBB}).

\begin{figure}[h!]
  \begin{center}
\includegraphics[width=8cm]{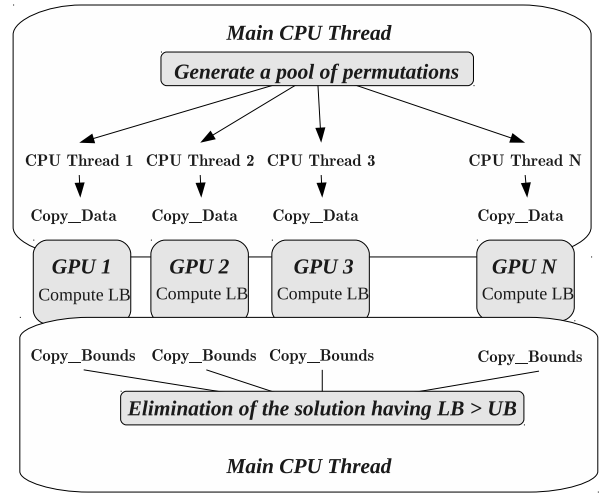}%
\caption{Parallel evaluation of bounds over multiple GPUs}
\label{multigpu}
  \end{center}
\end{figure}

As explained in Section \ref{GPUBB}, the main CPU thread selects a pool of unexplored nodes from the search tree according to their depth.
That pool of sub-problems is equally splitted into as much pools as the number of the used devices. 
In order to ensure complete concurrency between the bounding computations, we create as much CPU
threads\footnote{ We used lightweight threads defined by the POSIX Threads library.} as GPUs to be utilized. 
We assign to each thread CPU an individual GPU using the NVIDIA CUDA Runtime API ``cudaSetDevice()'' method \cite{NvidiaCudaGuide}, which gives the possibility to select which device to execute the kernel on. 
Each created thread CPU copies its pool of sub-problems from the CPU to its affiliated GPU, executes the kernel, and copies the resulting bounds 
back to the CPU. The main CPU thread waits for all other CPU threads to complete and merges results into one. The process is illustrated in figure~\ref{multigpu}. 
 
\section{Experiments}
\label{Experiments} 

In the following, an experimental study is presented with the objective to evaluate the performance impact of 
the presented auto-tuned GPU-accelerated B\&B algorithm and the effect of exploiting
larger degrees of parallelism by using multiple GPUs accelerators.

\subsection{Flow-shop instances} 

In our experiments, we used the flow-shop instances defined by Taillard \cite{Taillard_Bench}.
These standard instances are often used in the literature to evaluate the performance of algorithms that minimize the makespan.
Optimal solutions of some of these instances are still not known.
The different instances are designated by $n \times m$, where $n$ and $m$ represent respectively the number of 
jobs (between $20$ and $500$) to be scheduled and the number of machines ($20$,$10$,$5$) to be used. 
In our experiments, we used only the instances with $10$ or $20$ machines since instances with $5$ machines are easy to solve. For these instances, with $5$ machines, 
the used bounding operator gives so good lower bounds that it is posssible to solve them in few minutes using a sequential B\&B.
Therefore, these instances do not require the use of a GPU. We also omit instances with $500$ jobs because they do not fit in the memory of the CPU.

\subsection{Hardware and software platforms} 

The approach has been implemented using C-CUDA 4.0. The experiments have been carried out using 
an Intel Xeon E5520 bi-processor. This bi-processor is 64-bit, quad-core and has a clock 
speed of 2.27GHz. It is coupled with an Nvidia Tesla S1070 Computing System which is an 1U rack-mount
system equiped with two Tesla T10 GPUs. Each GPU contains 240 CUDA cores, 
a 4GB global memory, a 16.38KB shared memory, and a warp size of 32 threads. 
Using the occupancy calculator tool provided by NVIDIA, which allows the programmer to easily calculate 
the best thread block size based on register and shared memory usage of a kernel, we figure out that a block 
size equal to $256$ gives the best results. Therefore, we fixed the block size to $256$ in all our experiments. 
Here we notice that the number of threads per block is a multiple of the warp size 
\footnotetext{A warp contains 32 threads in the G80 model} which makes the kernel avoid
wasting computation on underpopulated warps. We vary the number of blocks in order to guarentee that the total number of active threads equals 
the size of the pool to submit to the GPU. As explained in Section \ref{GPUBB}, we first initialize the number of blocks with the number of the multiporcessors detected on the device. 

\subsection{Experimental protocol: speedup computation}

To evaluate the performance of the proposed approach, we calculate the speedup 
obtained by comparing our GPU B\&B version to a sequential B\&B version deployed on a single CPU core.
Since the used instances are very hard to solve (optimal solutions for many of these instances are still not known), 
we used the approach defined in \cite{Mezmaz_2007} to run experiments. Employing this method 
allows to obtain a random list $L$ of subproblems such as the resolution of $L$ lasts $Tcpu$ minutes with a sequential B\&B. 
To ensure that the subproblems explored by the GPU and CPU B\&B versions are exactly the same, we initialize the pool 
of our GPU-based B\&B with the same list $L$ of subproblems used in the sequential version. 
If we suppose the resolution of the GPU-based B\&B last $T{gpu}$ minutes, the reported speedup of our algorithm will be equal to $Tcpu/Tgpu$. 

\subsection{Performance impact of GPU-based parallelism}

The objective of the experimental study presented in this section is to demonstrate that the use of a GPU 
allows to significantly accelerate the execution time of the B\&B algorithm whatever is the FSP instance. The second objective 
is to find for each problem instance the best pool size that allows to take the most benefit from the use of the GPU.

\begin{figure}[h!]
  \begin{center}
\includegraphics[scale=0.6]{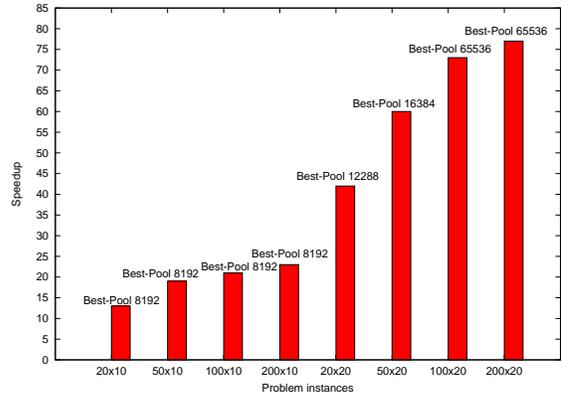}%
\caption{The speedups and corresponding used pools obtained using the auto-tuned algorithm.}
\label{speedupadatatif}
  \end{center}
\end{figure}

Figure~\ref{speedupadatatif} depicts the speedups obtained for the different problem instances using the 
approach proposed in Section \ref{GPUBB}. For each problem instance we report the best pool returned by our
dynamic parameter tuning heuristic. The reported results show that evaluating in parallel the bounds of 
a judiciously selected pool allows to significantly speedup the execution of the B\&B. Indeed, an acceleration factor
up to ($\times$78) is obtained for the 200 $\times$ 20 problem instances. The results show also that the parallel 
speedup grows with the size of the problem instance. For a fixed number of machines, the
obtained speedup grows accordingly with the number of jobs. For instance, the speedup obtained with 200 jobs ($\times$78) is
higher than the one obtained with 100 jobs ($\times$73), 50 jobs ($\times$62) and 20 jobs ($\times$44). This is due to the complexity of the computation of the 
lower bound which is $O(m^2.n.logn)$. For large problem instances (i.e. large values of $n$ and $m$) 
the grain size of the kernel executed by each thread is much higher which increases the GPU throughput.

To validate the proposed heuristic for auto-tuning the pool size, we run several experiments using different pre-fixed pool sizes.
The corresponding results are reported in Table~\ref{adaptatif}. The rows correspond to the problem instances defined by (Number of jobs $\times$ Number of machines)
and the columns correspond to the size of the pool of sub-problems to be evaluated in parallel.

Reported results clearly confirm that the best size of the pool depends strongly on the problem instance being solved. For instance, the best speedups for the 200 $\times$ 20 
instances are obtained with a pool size of 65536. However, with the 50 $\times$ 20 instances, the best speedup 
is obtained with a pool of 16384 problems. Another important result is that the best speedups measured when varying the sizes of the 
pool are obtained with the same pool sizes returned by our heuristic (see figure~\ref{speedupadatatif}). For example, the best speedup
for the 200 $\times$ 20 instances is obtained with a pool size of 65536 which is the best pool size our heuristic figure out for those instances.

\begin{table*}
\setlength{\tabcolsep}{0.3cm}
  \centering
  \footnotesize
 \begin{tabular}{|c|c|c|c|c|c|c|c|c|c}
    \hline
    \hline
(No. of jobs $\times$ No. of machines) &4096 & 6144 & 8192 & 12288 & 16384 & 32768 & 65536\\
    \hline
    \hline
$200 \times $20&40,49& 57,60 &62,64& 69,76&73,90&75,48&\textbf{78,14}\\
    \hline
$100 \times $20&42,79& 54,83 &61,96& 65,81 &69,93&71,57&\textbf{73,27}\\
    \hline
$50 \times $20&40,74& 51,40 &57,31& 60,89 &\textbf{62,15}&59,19&58,94\\
    \hline
$20 \times $20&33,38& 40,26 &43,60& \textbf{45,51} &44,16&39,75&38,36\\
    \hline
    \hline
$200 \times $10&19,34& 21,91 &\textbf{23,03}& 22,71 &22,11&21,68&21,09\\
    \hline
$100 \times $10&19,15& 20,76 &\textbf{22,09}& 21,72 &21,56&21,30&20,40\\
    \hline
$50 \times $10&18,21& 20,01 &\textbf{20,42}& 19,93 &19,55&18,77&18,25\\
    \hline
$20 \times $10&13,60& 14,81 &\textbf{15,03}& 14,58 &13,92&12,28&11,52\\
    \hline
    \hline
    \hline
 \end{tabular}
  \caption{Parallel speedup mesured for different problem instances and pool sizes without using the auto-tuning heuristic. }
\label{adaptatif}
\end{table*}

\subsection{Performance impact of MultiGPU-based parallelism}

In this section, we experiment the use of our parallel B\&B algorithm with
multiple GPUs. The objective here is to evaluate the impact of the multiGPU-based parallelism
proposed in section ~\ref{MULTIGPU}.

\begin{figure}[h!]
  \begin{center}
\includegraphics[width=9cm]{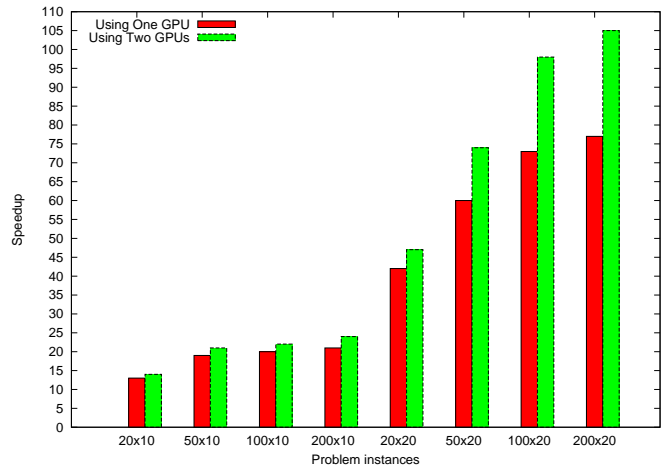}%
\caption{Comparing the parallel efficiency for different problem instances using a single / multiple GPUs.}
\label{speedupmultigpu}
  \end{center}
\end{figure}

Figure~\ref{speedupmultigpu} compares the computed speedups obtained for the different problem instances 
using respectiveley one and two GPU(s). The reported results show that evaluating bounds in parallel over two GPUs
provides further orders of speedups compared to an execution where only a single GPU is used whatever the instance is.
For instance, an acceleration factor up to $\times$105 is obtained with two GPUs for the 200 $\times$ 20 problem instances while
a speedup of $\times$78 is obtained for the same instances using only one device. In this case, executing the bounding operation on parallel GPUs 
provides an improvement about 26\% compared to a single GPU execution. 

The improvement we noticed when using two GPUs was somehow predictable. Indeed, exploiting the combined resources of multiple boards is 
promising for applications that have large amounts of parallelism. Our preliminary investigation \cite{Chakroun_2011} demonstrated that computing the lower bounds for the 
flow-shop permutation problem is one of those applications since significant acceleration factors have been mesured when running this function on parallel over a GPU.

\section{Conclusion and Future Work}
\label{Conclusion}

In this paper, we have presented new insights into parallel B\&B algorithms for solving permutation-based combinatorial optimization problems 
such as FSP on multiple GPU accelerators. The contributions consist in proposing an adaptative GPU-based parallel B\&B algorithms
and in exploiting larger degrees of parallelism by utilizing multiple GPUs. The Flow-Shop scheduling problem has been considered as a case study. The proposed approaches 
have been experimented on well-known FSP benchmarks using an Nvidia Tesla S1070 Computing System equipped with two Tesla T10 GPUs.

In our proposed auto-tuned GPU-based approach, the decomposition and pruning of the sub-problems is performed on CPU and the evaluation 
of their lower bounds (bounding operation) is executed on the GPU device. Pools of sub-problems are off-loaded from CPU to 
GPU to be evaluated by blocks of threads. After evaluation, the lower bounds are returned to the CPU. In order to dynamically tune
the size of the pool to be submitted to the GPU, we propose an heuristic for parameters auto-tuning at runtime. The main idea 
of this heuristic is to send the pending sub-problems using different-sized ``waves'' to the GPU device during the first iterations of the B\&B algorithm. 
Regularly, the efficiency of the used pool is computed and the size of the pool to off-load to the GPU is doubled. 
After a fixed number of trials, the best efficiency over all selected configurations is used for the remaining iterations of the algorithm.
The experimental results show that accelerations up to $\times 78$ can be obtained especially for large problem instances and large pools of sub-problems and
that the best speedups are obtained using the pool sizes returned by the heuristic. 

Another challenge of this work was to exploit larger degrees of parallelism by utilizing the combined computational power of multiple
graphical cards. Our concern towards a multi-GPU implementation of our parallel B\&B was to define a workload distribution between the used GPUs
in order to make all the available devices compute the same work in parallel without need of synchronization.
Experimental results demonstrate that using two GPUs is beneficial and improvement up to 23\% is reached compared to 
an execution with a single GPU. Thus, our proposed adaptative multi-GPU based Branch and Bound enables to achieve speedups of $\times$105 over a CPU version.

We are currently investigating the combination of the two parallel models Type~1 and Type~2 (see Section \ref{BBAlgo}) for the design and implementation of a GPU-accelerated 
multi-core B\&B algorithm. In the near future, we plan to extend this work to a cluster of GPU-accelerated multi-core processors. 
From application point of view, the objective is to solve to optimality challenging and unsolved Flow-Shop instances as 
we did it for one $50 \times 20$ problem instance with grid computing~\cite{Mezmaz_2007}. Finally, we plan to investigate other 
lower bound functions to deal with other combinatorial optimization problems.

\end{document}